\documentclass[twocolumn]{aastex631}

\begin{document}

\title{A Portrait of the Cosmic Reionisation History in the Context of the Early Dark Energy Model}

\correspondingauthor{Xin Wang, Hu Zhan}
\email{xwang@ucas.ac.cn \quad zhanhu@nao.cas.cn}


\author[0009-0004-9780-5979]{Weiyang Liu}
\affiliation{Key Laboratory of Space Astronomy and Technology, National Astronomical Observatories, Chinese Academy of Sciences,
20A Datun Road, Chaoyang District,
Beijing, 100101, China P.R.}
\affiliation{School of Astronomy and Space Science, University of Chinese Academy of Sciences,
1 Yanqihu East Rd, Huairou District,
Beijing, 101408, China P.R.}

\author[0000-0002-9373-3865]{Xin Wang}
\affiliation{Key Laboratory of Space Astronomy and Technology, National Astronomical Observatories, Chinese Academy of Sciences,
20A Datun Road, Chaoyang District,
Beijing, 100101, China P.R.}
\affiliation{School of Astronomy and Space Science, University of Chinese Academy of Sciences,
1 Yanqihu East Rd, Huairou District,
Beijing, 101408, China P.R.}
\affiliation{Institute for Frontiers in Astronomy and Astrophysics, Beĳing Normal University,
19 Xinjiekouwai Street, Haidian District,
Beijing, 100875, China P.R.}

\author[0000-0003-1718-6481]{Hu Zhan}
\affiliation{Key Laboratory of Space Astronomy and Technology, National Astronomical Observatories, Chinese Academy of Sciences,
20A Datun Road, Chaoyang District,
Beijing, 100101, China P.R.}
\affiliation{The Kavli Institute for Astronomy and Astrophysics, Peking University,
5 Yiheyuan Road, Haidian District,
Beijing, 100871, China P.R.}


\author[0000-0002-3254-9044]{Karl Glazebrook}
\affiliation{Centre for Astrophysics and Supercomputing, Swinburne University of Technology,
PO Box 218, Hawthorn,
VIC 3122, Australia}

\author[0000-0001-5940-338X]{Mengtao Tang}
\affiliation{Steward Observatory, University of Arizona, 933 N Cherry Ave, Tucson, AZ 85721, USA}

\author[0000-0001-9391-305X]{Michele Trenti}
\affiliation{School of Physics, University of Melbourne,
Parkville 3010,
VIC, Australia}
\affiliation{ARC Centre of Excellence for All Sky Astrophysics in 3 Dimensions (ASTRO 3D),
Melbourne,
Victoria, Australia}

\begin{abstract}


Recent JWST observations of Lyman-$\alpha$ emission at $z \sim 11–6$ indicate a rapid reionization of the intergalactic medium within the first $\sim700$ Myr. The required Lyman continuum (LyC) photon budget may naturally arise from the unexpectedly high galaxy number densities revealed by JWST, reducing the need for scenarios invoking very high LyC escape fractions ($f_{\rm esc}\gtrsim0.2$) or dominant contributions from ultra-faint galaxies ($M_{\rm UV}>-15$) in the standard $\Lambda$CDM framework. In this work, we model the reionization history under the Early Dark Energy (EDE) paradigm --- originally proposed to ease the Hubble tension --- which also explains the observed over-abundance of high-$z$ galaxies without extreme star formation efficiencies. The EDE model yields reionization histories consistent with current constraints while requiring only moderate LyC escape fractions and UV luminosity densities ($f_{\rm esc}\sim 0.05$–0.1, $M_{\rm UV}\lesssim -17$ to $-15$). Our results suggest that, once key astrophysical parameters are better constrained, the reionization history could serve as an independent and complementary probe of EDE cosmologies.

\end{abstract}



\section{Introduction} \label{sec:intro}

The driving source of cosmic reionisation has been a long-standing debate since the Epoch of reionisation was first identified by \cite{1965ApJ...142.1633G,2001AJ....122.2850B}. The joint observations of the Lyman-$\alpha$ Emitters \citep[LAEs, e.g.,][]{2024ApJ...975..208T,2025MNRAS.536.2355J,witstokWitnessingOnsetReionization2025}, Lyman Break Galaxies \citep[LBGs, e.g., ][]{2024ApJ...973....8H,2024ApJ...971..124U}, Lyman-$\alpha$ (Ly$\alpha$) and Lyman-$\beta$ (Ly$\beta$) forests \citep[e.g.,][]{2015MNRAS.447..499M,2023ApJ...942...59J}, Quasars \citep[e.g.,][]{2022MNRAS.512.5390G,2024ApJ...969..162D}, and Cosmic Microwave Background \citep[CMB,][]{2020A&A...641A...6P} indicate that the Universe is fully ionised within the period no longer than 700 Myr ($z\sim11-6$). However, the exact dynamics of reionisation remain veiled since the major source of the ionising photons has not yet been well identified. The current consensus inclines to the dwarf galaxies being the major provider of the ionising photons under the assumptions of a high escape fraction of the ionising photons that reionise the neutral Intergalactic Medium (IGM), $f_{esc}>0.2$ \citep[e.g.,][]{2025arXiv250105834K}, and a conservative production efficiency of the ionising photons, $\lg [\xi_{ion}/\text{Hz erg}^{-1}]\sim25.2$ \citep[see, e.g.,][]{2015ApJ...813...21M}.

However, the observations at $z\leq4$ indicate that the average $f_{esc}$ should not exceed $0.05$ except for those with fierce activities, such as Active Galactic Nuclei (AGNs) or gamma-ray bursts \citep[][]{2022ARA&A..60..121R,2025ApJ...980...74W,2003ApJ...598..878M,2020ApJ...904...59A,2016ApJ...831...38V,2015ApJ...804...17S,2010ApJ...723..241S,2007ApJ...668...62S,2003ApJ...598..878M}. Even though the faint galaxies at higher redshift are more likely to be bluer (in terms of the flux index $\beta)$ due to their younger stellar population and thus a higher $f_{esc}$ \citep[][]{2022MNRAS.517.5104C}, 
current observational evidence supporting a high average $f_{esc}>0.2$ remains limited. Instead, recent JWST observations, e.g., \cite{2025arXiv250705381Y,2025arXiv250508870P}, show that $f_{esc}$ for the majority of galaxies at $z>4.5$ is below $0.05$.

Meanwhile, the recent observational result of $\xi_{ion}$ from JWST implies that its value reaches up to $\lg [\xi_{ion}/\text{Hz erg}^{-1}]=25.8$ \citep[][]{2024Natur.626..975A}, approximately quadrupled that of the previous value. As mentioned in \cite{2019ApJ...879...36F}, a higher value of $\xi_{ion}$ is more likely to emerge in faint galaxies, due to their relatively younger stellar population and poorer metallicity. Hence, in the dwarf galaxy-dominated (i.e., democratic) model, the rapid completion of the reionisation process might be partly attributed to its efficient ionising photon production rate.

On top of that, the extent to which level of faintness should the dwarf galaxies be incorporated is yet an undetermined matter due to the scarcity of robust data above the Ultraviolet (UV) absolute magnitude $M_{UV}\gtrsim-17$ at $z\geq6$ \citep[e.g.,][]{2024Natur.633..318C,2023Natur.618..480R}. In order to produce sufficient ionising photons, a common approach is to include fainter galaxies down to $M_{UV}=-17$ to $-12$ \citep[e.g.,][]{2015ApJ...813...21M}, thereby offset the ionising photon deficit due to lower $f_{esc}$ or $\xi_{ion}$. On the other hand, the high redshift observations from JWST reveal surprisingly higher galaxy abundance than previously anticipated \citep[e.g.,][]{2023Natur.616..266L}, which resonates with the discovery of the massive quiescent galaxies, e.g., ZF-UDS-7329 \citep[][]{2024Natur.628..277G}, that implies extremely early formation time of their progenitors. 
In this case, one may expect that the swift reionisation history evolution is driven by the extra galaxies with moderate conditions of $f_{esc}$ or $M_{UV, lim}$. However, it raises another question: why were these additional galaxies not anticipated by previous predictions based on the $\Lambda$CDM model?

In order to answer this question, a number of unconventional galaxy formation models have been proposed to rationalise the appearance of these extra galaxies under the $\Lambda$CDM regime, such as the feedback-free starburst \citep[e.g.,][]{2025A&A...695A..20S}, younger stellar population \citep[e.g.,][]{2024MNRAS.533.3222D} and high star formation efficiency \citep[e.g.,][]{2023NatAs...7..731B,2025MNRAS.tmp.1706S}, 
Alternatively, the non-standard cosmology model of Early Dark Energy was also proposed to describe the evolution history of the galaxies \citep[][]{2024MNRAS.533..860L,2024MNRAS.533.3923S,2025PhRvD.111b3519J}, in which case the extra galaxies originate from the overall increment of the matter power spectrum rather than the modification of the individual galaxy formation model.

The EDE model \citep[][]{2016PhRvD..94j3523K,2018PhRvD..98h3525P,2019PhRvL.122v1301P,2023ARNPS..73..153K} was initially proposed to resolve the Hubble tension by introducing an extra substance in pre-recombination era to accelerate the cooling rate of the Universe, which leads to the increment of both $H_{0}$ and $\sigma_{8}$. The analysis of the CMB observations from the Atacama Cosmology Telescope (ACT) and South Pole Telescope (SPT) implies that EDE is a better fit of the data over $\Lambda$CDM \citep[][]{2021PhRvD.104l3550P,2022PhRvD.106d3526S} and generates a larger $H_{0}$ consistent with the local observations. However, these works, as well as the galaxy survey data, e.g., Dark Energy Survey (DES) or Baryon Oscillation Spectroscopic Survey (BOSS), also point out that no clear evidence of EDE emerges at the small scale ($\ell>1300$) Planck \citep[][]{2020A&A...641A...6P} or low redshift Large Scale Structure (LSS) observations, such as Baryon Acoustic Oscillation (BAO) or Redshift Distortion (RSD) \cite[][]{2020PhRvD.102d3507H}. Recently, the analysis of the Dark Energy Spectroscopic Instrument Data Release 2 (DESI DR2) data conducted by \cite{2025arXiv250324343C} suggests that EDE modestly improves the agreement between CMB and BAO with a $5$ per cent increment of $H_{0}$, which posits a potential positive argument for the model in the realm of LSS. In order to further test the performance of EDE in \textit{small scale} and \textit{low redshift}, \cite{2024MNRAS.533..860L} predicted the corresponding galaxy abundance and concluded that the EDE fits better with the LF data mainly at the bright end and high redshift ($z\geq10$). In this work, we aim to further check the credibility of the EDE model at the faint end with $z>6$ through the probability of providing a neat and coherent portrait of the neutral fraction evolution of IGM with mitigated dependence of the escape fraction or extremely faint galaxies, and evaluate to what extent the CMB-constrained EDE model can be probed by reionisation history.

The structure of this paper is as follows. In Section \ref{sec:theory} we present the details of the EDE theory and discuss the critical parameters involved in the calculation of the reionisation history. The results of the comparison between the prediction and observations are described in Section \ref{sec:results}. Finally, we summarise this work in Section \ref{sec:conclusion}. Throughout this paper, we apply the AB magnitude system \citep[][]{1983ApJ...266..713O}.

\section{theory} \label{sec:theory}
\subsection{Early Dark Energy Model} \label{sec:ede}
The need for the Early Dark Energy model arises from the recognition of the Hubble tension, and the speculation that the statistical or observational flukes alone bear limited capability to resolve it. Instead, the EDE model, an ad hoc postulate of a type of theoretical solution, suggests that the tension could be alleviated by introducing an extra substance into the pre-recombination epoch, through which the value of $H_{0}$ can be elevated due to a consequential faster cooling rate of the baryons. 

The CMB observations precisely measured the angular size of the sound horizon, $\theta_{s}\equiv r_{s}/D_{A}$, at the epoch of recombination, where $r_{s}$ and $D_{A}$ are the Euclidean size of the sound horizon and the distance towards the surface of the last scattering, respectively. For a given cosmological model, the corresponding $H_{0}$ can be derived as \citep[][]{2023ARNPS..73..153K}
\begin{equation}
    H_{0} = \sqrt{3}H_{ls}\theta_{s} \frac{\int_{0}^{\infty} dz [\rho(z)/\rho_{0}]^{-1/2}}{\int_{z_{ls}}^{\infty} dz[\rho(z)/\rho(z_{ls})]^{-1/2} (1+R)^{-1/2}}.
    \label{eq:h_cmb}
\end{equation}
Here $H_{ls}$ is the Hubble parameter at the last scattering surface extrapolated from the cosmological parameters at $z=0$, $\rho(z)$ and $\rho_{0}$ are the total energy density at redshift $z$ and present, respectively, and $R$ represents the density ratio of the baryons to photons, $R=(3/4)(\omega_{b}/\omega_{\gamma})/(1+z)$. Therefore, the value of $H_{0}$ can be elevated by increasing the energy density $\rho(z)$.

In this paper, we focus on the \textit{Axion-like} EDE model \citep[][]{2016PhRvD..94j3523K,2018PhRvD..98h3525P,2019PhRvL.122v1301P,2023ARNPS..73..153K}, in which it behaves as an oscillating scalar field $\phi$, 
\begin{equation}
    V_n(\phi) = m^{2}f^{2}[1 - \cos{(\theta)}]^{n},
    \label{eq:v_n}
\end{equation}
where $V_{n}(\phi)$ is the potential of the field, with index $n$ controls its decay rate through the equation of state, $w_{n}=(n-1)/(n+1)$. The canonical axion dark matter model and radiation are described by $n=1$ and $2$, respectively, and the EDE model has $n>2$, which results in a larger decay rate than radiation. For instance, $\rho_{EDE}(z)\propto a^{-9/2}$ when $n=3$, the best-fit value according to current CMB data \citep[][]{2020PhRvD.101f3523S,2022PhRvD.106d3526S,2021PhRvD.104l3550P}, and almost entirely dissipated after recombination \citep[Figure 2 in][]{2025arXiv250324343C}. The $m$ is the effective mass of the axion particle, and $f$ represents the decay constant. $\theta\equiv\phi/f$ is the re-normalisation factor such that $-\pi\leq\theta\leq\pi$. The EDE parameters applied for this work are listed in Table \ref{tab:ede_params}. Specifically, we choose the values fitted from pure CMB observations of ACT+SPT+Planck from \cite{2022PhRvD.106d3526S} to avoid possible bias from the low-redshift prior.

\begin{table}
    \centering
    \begin{tabular}{lc}
        \hline
        \hline
        Parameter       & EDE       \\
        \hline
        $h$             & 0.7420    \\
        $\omega_{c}$    & 0.1356    \\
        $n_{s}$         & 1.0010    \\
        $\sigma_{8}$    & 0.8446    \\
        $n$             & 3         \\
        $\lg({z_{c}})$  & 3.526     \\
        $f_{c}$         & 0.163     \\
        \hline
    \end{tabular}
    \caption{The cosmological parameters adopted for the EDE models. We take the best-fit values based on the CMB observations, i.e., ACT+SPT+Planck TT650TEEE, from \citet{2022PhRvD.106d3526S}.}
    \label{tab:ede_params}
\end{table}

\subsection{Reionisation History} \label{sec:reio_his}
The evolution of the reionisation history is concisely described by the ionised fraction of the hydrogen in IGM, $Q(z)$, in terms of the competition between the rates of ionisation and recombination of the initially neutral hydrogen atoms \cite[][]{1999ApJ...514..648M},
\begin{equation}
    \frac{dQ(z)}{dt} = \frac{\dot{n}_{ion}}{\langle n_{H}\rangle} - \frac{Q(z)}{\bar{t}_{rec}}.
    \label{eq:dqdt}
\end{equation}
Here $\dot{n}_{ion}$ is the comoving number density of the ionising photons produced by the reionisation source galaxies, 
\begin{equation}
    \dot{n}_{ion} = f_{esc}\xi_{ion}\rho_{UV},
    \label{eq:n_ion}
\end{equation}
where $f_{esc}$ represents the escape fraction of the photons from their sources to ionise IGM, $\xi_{ion}$ is the production rate of the ionisation photons in the unit of Hz erg$^{-1}$, and $\rho_{UV}$ is the luminosity density in the unit of erg s$^{-1}$ Hz$^{-1}$ Mpc$^{-3}$, which quantifies the integrated energy density of the ionisation source galaxies above a given luminosity threshold (i.e., $M_{UV,lim}$). $\langle n_{H}\rangle$ is the mean of the hydrogen atoms comoving  density, and $\bar{t}_{rec}$ is the average recombination timescale of the IGM \citep[][]{2004ApJ...600..508S,2015ApJ...802L..19R}, 
\begin{equation}
    \bar{t}_{rec}(z) = [C\alpha_{B}(T)n_{e}(1+z)^3]^{-1}.
    \label{eq:t_rec}
\end{equation}
The factor $C=\langle n_{H}^{2} \rangle / \langle n_{H} \rangle^{2}$ is the clumping factor of the hydrogen for which we use a uniform distribution between $1$ and $6$, $U(1, 6)$, following the approach of \cite{2015ApJ...813...21M}. $\alpha_{B}$ is the case B recombination coefficient that takes the absorption of the photons by the opaque IGM into account. The density of the electron is given by $n_{e}=(1+Y_{p}/4X_{p})/\langle n_{H} \rangle$, where $X_{p}$ and $Y_{p}$ are the mass fractions of hydrogen and helium, respectively. We list the values of these parameters applied in this work in Table \ref{tab:reio_params}.
\begin{table*}
    \centering
    \begin{tabular}{llll}
        \hline
        \hline
        Parameter       & Value                 & Unit                      & Note                                                          \\
        \hline
        $f_{esc}$       & $0.05, 0.1, 0.2$      &  -                        & -                                                             \\
        $\lg\xi_{ion}$  & $25.8, 25.3$          &  Hz erg$^{-1}$            & \cite{2024Natur.626..975A,2024MNRAS.535.2998S}        \\
        $\rho_{UV}$     & $M_{UV, lim} = -12, -15, -17$       &  erg s$^{-1}$ Hz$^{-1}$ Mpc$^{-3}$     
        & $\rho_{UV}$ defined in Equation \ref{eq:rhouv}.                                                                   \\
        $C$             & $U(1, 6)$             & -                         & Clumping factor                                               \\
        $\alpha_{B}$    & $2\times10^{-13}$     & cm$^{3}$ s$^{-1}$         & Assume $T_{IGM}=20000$ K                                      \\
        $n_{H}$         & $1.94\times10^{-7}$   & cm$^{-3}$                 & $\sim2\%$ higher in EDE, no significant effect to $Q(z)$  \\
        \hline
    \end{tabular}
    \caption{The values of parameters applied for reionisation history (Equation \ref{eq:dqdt}).}
    \label{tab:reio_params}
\end{table*}

The main challenge to determine the reionisation history arises from the fact that the production rate of the ionising photons (Equation \ref{eq:n_ion}) depends on the galaxies at $z\geq6$, the properties of which could intrinsically differ from their local counterparts (e.g., the star-forming galaxies) and remain observational unconstrained as yet, due to the current observational limits. In the following sections, we elaborate on the fiducial values of these parameters for this work and the corresponding argumentations for a robust reconstruction of the reionisation history under the EDE model.

\subsubsection{Lyman Continuum Photon Escape Fraction} \label{sec:f_esc}
The direct constraints of the Lyman continuum photon escape fraction at $z>4$ is observationally inaccessible due to the opacity of the foreground IGM \citep[][]{2022ARA&A..60..121R}. Similar to the studies of the high-redshift galaxies, the current results of $f_{esc}$ at $z>6$ are mainly the analogues of the galaxies at lower redshift. In general, there are six factors that may affect the observed $f_{esc}$: redshift, observation effects (viewing angle and geometry of the galaxy), contamination of the non-ionising photons, foreground IGM absorption, presence of AGN, and the state of the galaxy itself (age, metallicity, stellar population, SFR, dust, feedback and outflow). As a result, its value could vary in a wide range from $f_{esc}\leq0.05$ to $f_{esc}\sim0.5$ \citep[][and references therein]{2022ARA&A..60..121R}.


Through the analysis of the indirect probes, \cite{2015ApJ...810..104A,2018ApJ...869..123S} found that $f_{esc}$ is strongly correlated with the equivalent width of Ly$\alpha$, and \cite{2017A&A...605A..67C} suggested that a strong outflow may create low $\rm HI$ density and metallicity channels for the escape of the ionising photons. This is also reflected in the empirical relation between $f_{esc}$ and the UV spectral slope $\beta$, a lower value of which represents a bluer flux that implies a young, metal-poor, and low dust attenuation galaxy. For instance, a recent result from \cite{2025arXiv250103217D} found that $f_{esc}\sim0.07$ and $0.19$ for $\beta=-2.22$ and $-2.6$, respectively. As mentioned in \cite{2022ARA&A..60..121R}, the empirical relations are established from the low-redshift observations, which indicate that the younger and bluer galaxies, or those with acute activities, e.g., vigorous star formation, strong line-emitting, AGN, or gamma-ray burst, bear high $f_{esc}\gtrsim0.2$, while the majority of the normal galaxies, due to their relatively tranquil dynamics, perform much lower escape fractions $f_{esc}\leq0.05$ \citep[e.g.,][]{2025ApJ...980...74W}. On top of that, the simulations also show that the feedback allows the $f_{esc}$ be boosted up to $\gtrsim0.2$ for the vigorous star-forming galaxies and, inversely, the escape fraction decreases for the massive and low mass objects due to their dusty environment or inefficiency in creating feedback channels for the ionising photons to escape, respectively \citep[][]{2020MNRAS.498.2001M}. 

Observationally, the results from JWST, e.g., \cite{2024A&A...685A...3M,2025arXiv250705381Y,2025arXiv250508870P}, indicate that the average $f_{esc}$ during EoR should not exceed $\sim0.1$. Therefore, although the galaxies at $z\geq6$ are expected to be younger and more active than their local counterparts, hence allowing a higher average $f_{esc}$ to take place, the observational evidence is still limited in supporting a high $f_{esc}>0.2$ for the reionisation source galaxies. Bearing this in mind, we take $f_{esc}=0.05, 0.1, 0.2$ in our calculation.

\subsubsection{Production Rate of LyC Photons}  \label{sec:xi_ion}
The value of the LyC photon production efficiency $\xi_{ion}$ is mainly indirectly measured by UV spectra, such as Ly$\alpha$, $\rm CIII]$ and $\rm [OIII$]+$\rm H\beta$ \citep[][]{2022ARA&A..60..121R}. Unlike $f_{esc}$, the measurement of $\xi_{ion}$ at $z\geq6$ is feasible. For instance, \cite{2024Natur.626..975A}  analysed 8 JWST sources spanning the redshift of $z\approx6-7.7$ and found that $\lg[\xi_{ion}/Hz \ erg^{-1}]=25.80\pm0.14$ for galaxies with $M_{UV}=-17$ to $-15$, which is higher than the canonical value of $\lg[\xi_{ion}/Hz \ erg^{-1}]=25.20$ \citep[][]{2015ApJ...802L..19R}. 

In general, galaxies with low metallicity and high binary interactions exhibit higher $\xi_{ion}$ due to their efficient fuel consumption. As a result, the application of Binary Population and Spectral Synthesis \citep[BPASS,][]{2017PASA...34...58E} IMF shows higher $\xi_{ion}$ than Salpeter IMF \citep[][]{1955ApJ...121..161S} for the former includes the involvement of binary systems. On top of that, according to the analysis of the nebular emission lines, bluer and strong-emission galaxies also have higher $\lg\xi_{ion}$. In light of the assumption that the reionisation-sourcing galaxies are younger galaxies with lower metallicity and higher star formation rate, we set $\lg[\xi_{ion}/Hz \ erg^{-1}]=25.80$ as the benchmark value for our calculation, and consider another result from JWST aligning with previous observations, $\lg[\xi_{ion}/Hz \ erg^{-1}]=25.30$ \citep[][]{2024MNRAS.535.2998S}, for comparison.

\subsubsection{UV Luminosity Density}   \label{sec:rho_uv}
The UV luminosity function $\rho_{UV}$ is defined by the integral of the luminosity function $\Phi_{UV}$, d
\begin{equation}
    \rho_{UV} = \int_{-\infty}^{M_{UV, lim}} \Phi_{UV} M_{UV} dM_{UV},
    \label{eq:rhouv}
\end{equation}
which represents the total energy emitted from the galaxies down to a given limit, $M_{UV, lim}$. Specifically, the rest-frame UV band is widely used for the observations of high-redshift galaxies since it is tightly connected to the formation rate of the young and massive stars. 

\textbf{Luminosity Function} is the most direct measure of the galaxy abundance. However, its theoretical derivation requires a roundabout process through the scaling of the Halo Mass Function (HMF), i.e., 
\begin{equation}
    \Phi_{UV} = \Phi_{h} \frac{dM_{h}}{dM_{\ast}} \frac{dM_{\ast}}{dM_{UV}}.
    \label{eq:phiuv}
\end{equation}
In this paper, we adopt the same method to derive $\Phi_{UV}$ as in \cite{2024MNRAS.533..860L}, with the parameters for the EDE model listed in Table \ref{tab:ede_params}. Specifically, we note that the JWST-based Star Formation Efficiency (SFE, i.e., $\epsilon$) and $M_{\ast}-M_{UV}$ relation could considerably differ from the previous empirical relations. For instance, \cite{2025arXiv250103217D} attempted to explain the JWST results by applying a younger stellar population and found higher $\epsilon$ for the low mass haloes than the pre-JWST results \citep[e.g.,][which we adopt for this work]{2021ApJ...922...29S}. Owing to the complexity of the galaxy formation theory, most of the current SFEs are not directly measured, but empirically fitted to match the existing data. Meanwhile, the derivation of it \citep[e.g.,][]{2025MNRAS.tmp.1706S} depends on a number of idealised assumptions. Therefore, the SFE from the pre-JWST observations may remain valid, should the overabundance of the galaxy from JWST be primarily driven by other factors, e.g., the alteration of cosmology.

The application of the EDE model itself does not change the fundamental properties of dark matter. However, we notice that the formation and evolution histories of the dark matter haloes and their corresponding galaxies may still deviate in different cosmologies due to the alteration of the initial conditions. Since the detailed conclusions regarding this matter require high-resolution simulations under the EDE model, which were not available when this paper was written, we hypothesise that the scaling relation of $M_{h}-M_{\ast}$, particularly for low mass systems, is insensitive to the cosmological variations, the reasons for which are described as follows.

The EDE theory regulates the galaxy number density through its higher $\sigma_8$ (Table \ref{tab:ede_params}). Consequently, the initial matter density fluctuations are easier to collapse and form the seed haloes at an earlier stage of the Universe. In this case, the concentration parameter \citep[$c$ in Navarro-Frenk-White profile, hereafter NFW, ][]{1997ApJ...490..493N} is expected to be larger than in the $\Lambda$CDM model \citep[][]{2009ApJ...707..354Z}. Given this, a naive extrapolation would be that a longer timescale with higher efficiency is given to the star formation, hence $M_{\ast}/M_{h}$ should be escalated under the EDE framework.

However, the initiation of the star formation requires the injected hydrogen molecular gas to cool as low as $\sim100$ K. This becomes particularly difficult for the first stars given earlier formation time because 1) the highly inefficient cooling process of the hydrogen atoms \citep[][]{2023ARA&A..61...65K}; 2) a denser and hotter environment increases the virial temperature ($T_{vir}\propto M_{h}^{2/3} \rho_{m}^{1/3}$) and suppresses the formation of hydrogen molecule; 3) stronger cosmological Lyman-Werner (LW) background radiation to photodissociate $H_{2}$ and jeopardise the follow-up star formation \citep[][]{2023ARA&A..61...65K}; 4) in case the previous 3 factors do not significantly suppress the formation of the first stars, an earlier appearance of the Supernovae (SNe) and AGN may take place that exerts preemptive feedback process as well. In turn, the earlier assembly and longer evolution time of the dark matter haloes do not necessarily lead to a more efficient channel for the proliferation of the stellar mass. Instead, the collective result could as well be the cancellation of the two sides of the galaxy formation process and, consequently, a cosmological background-insensitive $M_{h}-M_\ast$ relation during the epoch of reionisation. On top of that, this effect could be particularly true for the low-mass galaxies due to their shallower gravitational potentials. In light of the cumbersome nature of this issue, as well as the fact that the modification of the scaling relation and the alteration of cosmology are two independent methods to explain the galaxy abundance, we therefore simplify our assumption by applying the pre-JWST $M_{h}-M_{\ast}$ relation rather than the JWST-modified case to avoid changing two variables simultaneously that causes a redundant double-fitting of the model, and leave the quantitative verification of this matter to the future EDE-based simulations, such that the model can be further verified or falsified by the quantification of the scaling relation.



\textbf{Faint galaxies inclusion} $M_{\text{UV,lim}}$ represents the extent to which the faint galaxies should be considered the source of reionisation. The faintest observations of the UV LF to date have reached $M_{UV}=-15$ \citep[][]{2016MNRAS.456.3194P,2018MNRAS.479.5184A}. However, considering the consensus on the contribution of the faint galaxies yet remains an open question due to the lack of solid observational, particularly spectroscopic, determination of LF above $M_{UV}=-17$ during EoR \citep[][]{2024Natur.633..318C}, we adopt the common choice of $M_{\text{UV,lim}}=-17, -15, -12$ for this work.

\section{Results and Discussion} \label{sec:results}
\subsection{Galaxy Abundance} \label{sec:results_gal_abun}
\begin{figure*}
    \centering
    \includegraphics[width=\linewidth]{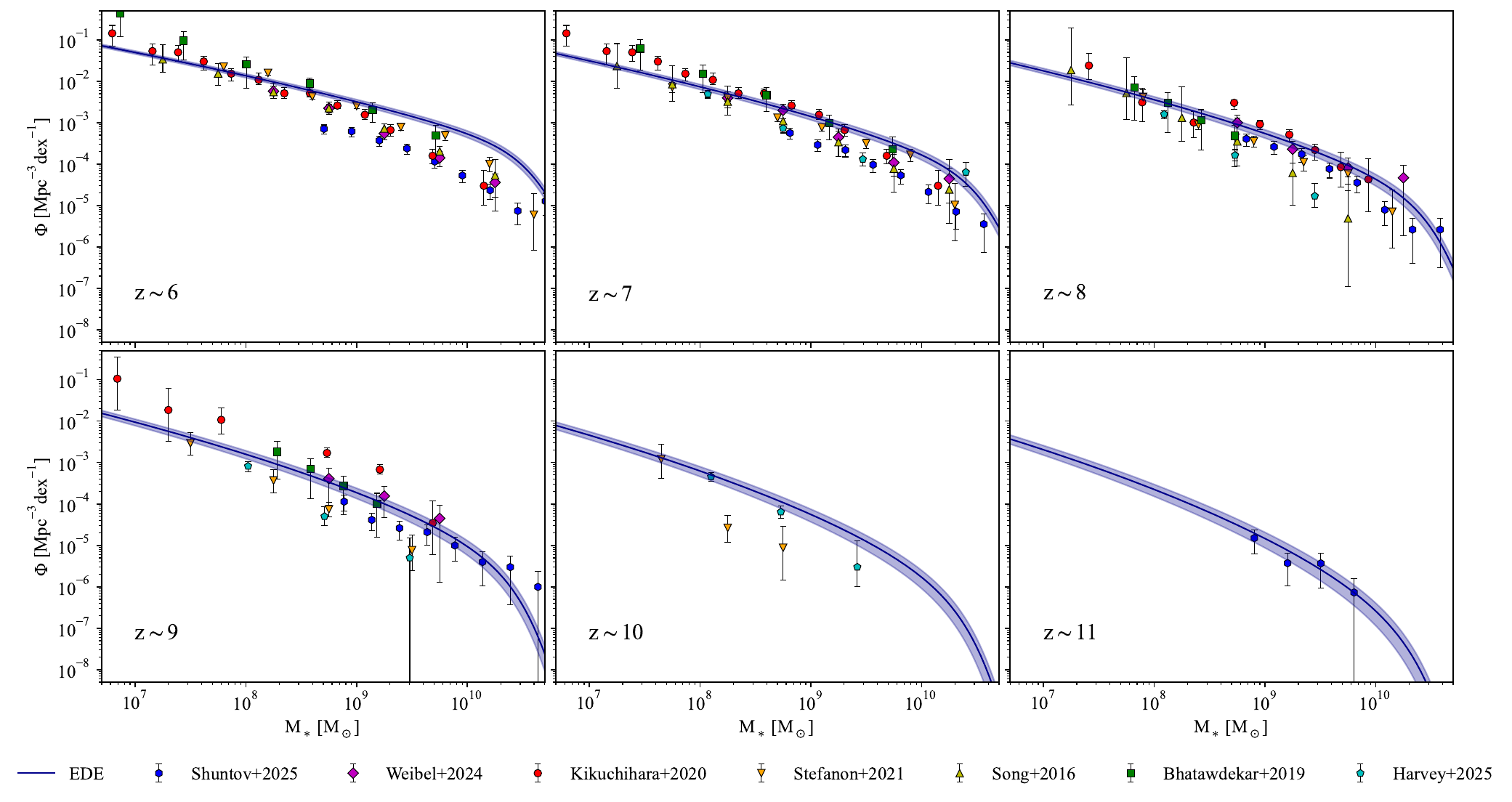}
    \caption{The prediction of SMF at redshift $z\sim6-11$ under 
    EDE model with the $M_h-M_\ast$ scaling relation from \citet[][]{2021ApJ...922...29S}. The error shade represents the uncertainties of HMF and the scaling relation. The data points are from both JWST \citep[][]{2025A&A...695A..20S,2024MNRAS.533.1808W,2025ApJ...978...89H} and pre-JWST \citep[][]{2021ApJ...922...29S,2016ApJ...825....5S,2020ApJ...893...60K,2019MNRAS.486.3805B}. The EDE model predicts relatively higher galaxy abundance at each redshift due to its elevated $\sigma_{8}$ (Table \ref{tab:ede_params}), and shows consistency with the observations at $z\sim7-11$. In particular, the prediction is consistent with the low mass end at each redshift bin, which is of particular importance for the reionisation.}
    \label{fig:smf}
\end{figure*}

Figure \ref{fig:smf} demonstrates the Stellar Mass Function (SMF) that bridges the HMF and LF through Equation \ref{eq:phiuv}, under the EDE model in the redshift $z\sim6-11$. The error shades are propagated from the uncertainties of the HMF and the $M_h-M_\ast$ scaling relation.
The data points are from both JWST \citep[][]{2025A&A...695A..20S,2024MNRAS.533.1808W,2025ApJ...978...89H} and pre-JWST \citep[][]{2021ApJ...922...29S,2016ApJ...825....5S,2020ApJ...893...60K,2019MNRAS.486.3805B}. 
As mentioned by \cite{2024MNRAS.533..860L}, the EDE-based galaxy abundance prediction, enhanced by the higher HMF caused by the escalated clustering of the matter distribution reflected in $\sigma_8$ (Table \ref{tab:ede_params}), is consistent with the high-redshift observations ($z\geq7$ in Figure \ref{fig:smf}). Notably, the faint end of the SMF is of particular importance for this work, as the reionisation is expected to be driven by the numerous low-mass galaxies. In this case, the EDE model provides a good fit for the entire redshift range from $z=6$ to $11$. We notice that there is a slight overestimation of the high mass end at $z\sim6$, and consider it might be attributed to the selection effects or the onset of the substantial AGN feedback that warrants further investigation.

\begin{figure*}
    \centering
    \includegraphics[width=\linewidth]{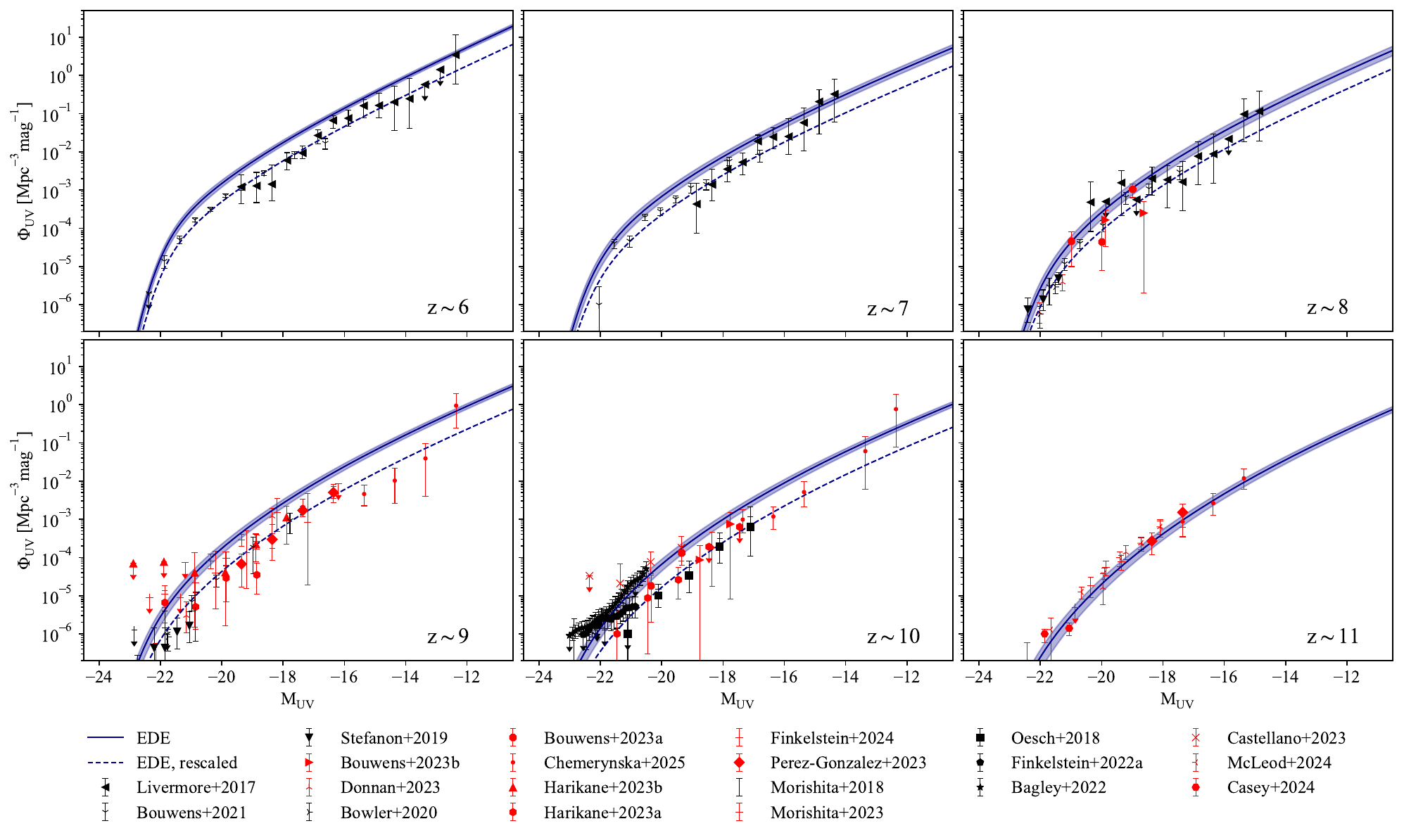}
    \caption{
    The LF at $z\sim6-11$ under the EDE model with the scaling relation from \cite{2021ApJ...922...29S}. The error shade incorporates the uncertainty from the $M_\ast-M_{UV}$ relation, in addition to those from SMF in Figure \ref{fig:smf}. The predictions from EDE are well agreed with the data at $z\sim7, 8, 11$ as well as the bright and faint ends of other redshifts. The overestimation in the intermediate brightness range, on the other hand, might be the result of the instantaneous SFR and dust attenuation, which could be verified with future observations. We also plot the rescaled LF at $z\sim6-8$ by $1/3$ and $z\sim8-10$ by $1/4$ to match the data points with intermediate brightness (dashed line), in which case the reionisation would be greatly delayed due to the insufficiency of the faint galaxies. We plot the data points from \citet{2023ApJ...948L..14C,2022ApJ...940L..55F,2023arXiv230406658H,2023ApJS..265....5H,2023ApJ...946L..35M,2022ApJ...940L..14N,2023ApJ...951L...1P,2023MNRAS.518.6011D,2023MNRAS.523.1009B,2023MNRAS.523.1036B,2022arXiv220512980B,2021AJ....162...47B,2022ApJ...928...52F,2018ApJ...867..150M,2018ApJ...855..105O,2020MNRAS.493.2059B,2019ApJ...883...99S,2023arXiv231104279F,2024MNRAS.527.5004M,2024ApJ...965...98C,2017ApJ...835..113L,2025arXiv250924881C}. The black data points are collected from pre-JWST observations, and the red data points are from JWST.
    }
    \label{fig:lf}
\end{figure*}

The luminosity density (Equation \ref{eq:rhouv}) is dominated by the faint end of the LF due to their higher number density. Figure \ref{fig:lf} shows the LF predicted by the EDE model as well as the observation data from \cite{2023ApJ...948L..14C,2022ApJ...940L..55F,2023arXiv230406658H,2023ApJS..265....5H,2023ApJ...946L..35M,2022ApJ...940L..14N,2023ApJ...951L...1P,2023MNRAS.518.6011D,2023MNRAS.523.1009B,2023MNRAS.523.1036B,2022arXiv220512980B,2021AJ....162...47B,2022ApJ...928...52F,2018ApJ...867..150M,2018ApJ...855..105O,2020MNRAS.493.2059B,2019ApJ...883...99S,2023arXiv231104279F,2024MNRAS.527.5004M,2024ApJ...965...98C,2017ApJ...835..113L,2025arXiv250924881C}. The error shade also takes the uncertainty from the $M_\ast-M_{UV}$ relation into account, in addition to those from SMF (Figure \ref{fig:smf}). We plot the pre-JWST results in black points and the JWST results in red. Roughly speaking, the EDE predictions are consistent with the observations at $z\sim7, 8$ and $11$, as well as the bright and faint ends for other redshifts. Whilst \cite{2024MNRAS.533..860L} proposed a suppression mechanism evolving with redshift to reconcile the discrepancy between the EDE prediction and observations at $z<10$ based on their data selection, some other observations from HST and JWST, e.g., \cite{2017ApJ...835..113L,2025arXiv250924881C} which are also plotted in Figure \ref{fig:lf}, detected higher faint-ends of LF, thus enhances the consistency between EDE prediction and observations. Further studies, such as \cite{2025arXiv250100984W}, also indicated that the faint end of the galaxy abundance may also exceed the previous anticipation. Bearing this in mind, we consider that the galaxy abundance with intermediate brightness at $z=6, 9$ and $10$ might be affected by the SFR at which they were observed or the dust attenuation, thus requiring further verification with extra observations. Alternatively, we manually rescale the LF at $z\sim6-8$ by $1/3$, and $z\sim8-10$ by $1/4$ to match the data points with intermediate brightness (dashed line). As shown below, our calculation of $Q(z)$ implies that the extra galaxies generated by the EDE model are necessary to produce sufficient ionising photons, for the insufficiency of the faint galaxies in the rescaled cases would otherwise greatly delay the entire reionisation process.

\subsection{Reionisation History under EDE}\label{sec:results_reio}
\begin{figure*}
    \centering
    \includegraphics[width=\linewidth]{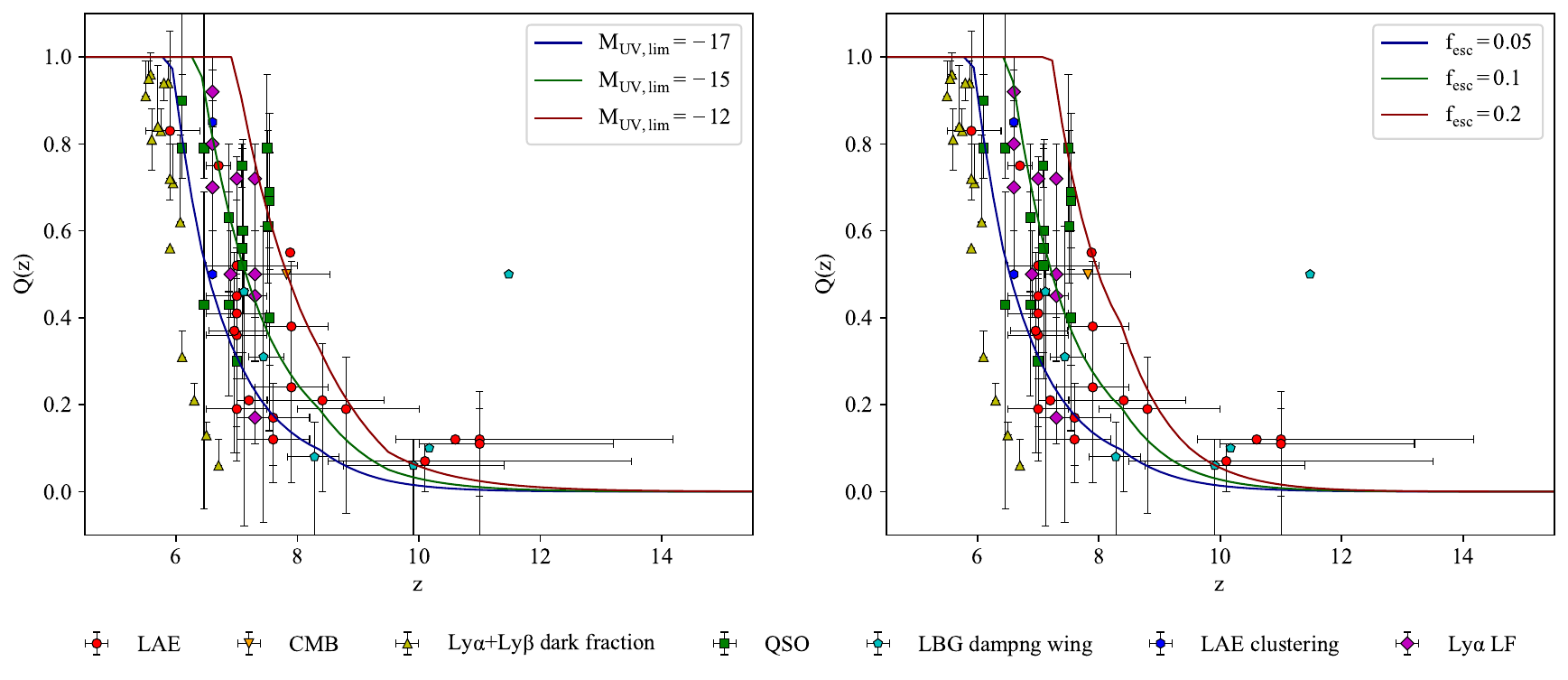}
    \caption{
    The comparison between reionisation history observations and EDE-based predictions with various parameter combinations. (Left) When adopting $f_{esc}=0.05$ and $\lg [\xi_{ion}/Hz \ erg^{-1}]=25.8$, a conservative choice of $M_{\text{UV,lim}}\leq-15$ provides a good fit to the data. (Right) With $M_{\text{UV,lim}}=-17$, a low value of $f_{esc}\sim0.1$ is sufficient for predicting a coherent evolution of the neutral fraction. In case the completion time of EoR is prioritised, the requirements of both parameters can be reduced to $f_{esc}\sim 0.05$ and $M_{UV, lim}\sim -17$.
    The data points are from various probes, i.e.,  Lyman-$\alpha$ Emitter \citep[][]{2018ApJ...856....2M,2019MNRAS.485.3947M,2019ApJ...878...12H,2020MNRAS.495.3602W,2022MNRAS.517.3263B,2023ApJ...949L..40B,2023ApJ...947L..24M,2024ApJ...967...28N,2025MNRAS.536.2355J,2024ApJ...975..208T,2025arXiv250105834K,2025MNRAS.538L..16Q}, CMB \citep[][]{2020A&A...641A...6P}, Lyman-$\alpha$ and Lyman-$\beta$ forest dark fraction \citep[][]{2015MNRAS.447..499M,2023ApJ...942...59J}, Quasars \citep[][]{2018Natur.553..473B,2018ApJ...864..142D,2020ApJ...896...23W,2020ApJ...897L..14Y,2022MNRAS.512.5390G,2024ApJ...969..162D}, Lyman-break galaxy \citep[][]{2023NatAs...7..622C,2024ApJ...973....8H,2024ApJ...971..124U}, Lyman-$\alpha$ Emitter clustering \citep[][]{2015MNRAS.453.1843S,2018PASJ...70S..13O}, and Lyman-$\alpha$ LF \citep[][]{2010ApJ...723..869O,2014ApJ...797...16K,2018PASJ...70S..16K,2017ApJ...842L..22Z,2018PASJ...70...55I,2021ApJ...923..229G,2021ApJ...919..120M,2022ApJ...926..230N}. 
    }
    \label{fig:q_test}
\end{figure*}

On the left side of Figure \ref{fig:q_test}, we show the evolution of $Q(z)$ in the EDE model for different values of $M_{UV,lim}$, with $\lg [\xi_{ion}/Hz \ erg^{-1}]=25.8$ and $f_{esc}=0.05$. The data points for comparison are from various probes, i.e., Lyman-$\alpha$ Emitter \citep[][]{2018ApJ...856....2M,2019MNRAS.485.3947M,2019ApJ...878...12H,2020MNRAS.495.3602W,2022MNRAS.517.3263B,2023ApJ...949L..40B,2023ApJ...947L..24M,2024ApJ...967...28N,2025MNRAS.536.2355J,2024ApJ...975..208T,2025arXiv250105834K,2025MNRAS.538L..16Q}, CMB \citep[][]{2020A&A...641A...6P}, Lyman-$\alpha$ and Lyman-$\beta$ forest dark fraction \citep[][]{2015MNRAS.447..499M,2023ApJ...942...59J}, quasars \citep[][]{2018Natur.553..473B,2018ApJ...864..142D,2020ApJ...896...23W,2020ApJ...897L..14Y,2022MNRAS.512.5390G,2024ApJ...969..162D}, Lyman-break galaxy \citep[][]{2023NatAs...7..622C,2024ApJ...973....8H,2024ApJ...971..124U}, Lyman-$\alpha$ Emitter clustering \citep[][]{2015MNRAS.453.1843S,2018PASJ...70S..13O}, and Lyman-$\alpha$ LF \cite[][]{2010ApJ...723..869O,2014ApJ...797...16K,2018PASJ...70S..16K,2017ApJ...842L..22Z,2018PASJ...70...55I,2021ApJ...923..229G,2021ApJ...919..120M,2022ApJ...926..230N}. The $\chi^2$ analysis results of each parameter set are listed in Table \ref{tab:ede_chi2}.

\begin{table}
    \centering
    \begin{tabular}{lcc}
        \hline
        \hline
        Parameter               & Value     & $\chi^2$      \\
        \hline
        $f_{esc}$               & 0.05      & 26.00         \\
        ($M_{UV, lim}=-17$)     & 0.1       & 11.50         \\
                                & 0.2       & 14.58         \\
        \hline
        $M_{UV, lim}$           & -17       & 25.89         \\
        ($f_{esc}=0.05$)        & -15       & 10.09         \\
                                & -12       & 11.90         \\
        \hline
    \end{tabular}
    \caption{The $\chi^2$ for various choices of $f_{esc}$ and $M_{UV, lim}$ under the EDE model. The results show that the combination of $(f_{esc}, M_{UV, lim})=(0.1, -17)$ and $(0.05, -15)$ provides the lowest $\chi^{2}$, implying that the EDE model alleviates the requirements of the two astrophysical parameters.}
    \label{tab:ede_chi2}
\end{table}

The contribution of the faint galaxies is a long-standing debate over the issue of reionisation. Given $f_{esc}=0.05$, the results of the curves and $\chi^2$ analysis suggest that $M_{\text{UV,lim}}\sim-15$ corresponds to a more consistent prediction with the data. On top of that, if the completion time of EoR is prioritised, the upper limit of $M_{UV}$ can be escalated to $-17$. 
In this case, a clear turnover of the luminosity function at $M_{UV}>-15$ should appear due to the hypothetical inefficient galaxy formation. Although a recent work from \cite{2025arXiv250924881C} reports no turnover in the faint-end LF up to $M_{UV}\sim-12.5$, the statistical significance of this result could still be limited by small-number statistics. Hence, the exact location at which the turnover takes place remains observationally unconstrained and requires deeper and richer accumulation of observation data.

On the other hand, we notice that the value of $\lg \xi_{ion}$ applied for this work is considerably higher than the canonical value. In light of some recent studies reporting lower values of it which is similar to the pre-JWST result \citep[e.g.,][]{2025MNRAS.537.3245B,2025ApJ...981..134P,2024MNRAS.535.2998S}, a fraction of fainter galaxies with $M_{UV}=-15$ to $-12$ or higher $f_{esc}$ (e.g., Figure \ref{fig:q_test}) may still be required to compensate for the reduction of the ionising photon supply from individual galaxies, should $\lg \xi_{ion}$ be confirmed not as high as we apply for this work. Therefore, the observational confirmation of $M_{UV}$ at which the LF turnover takes place, and better constraints on $\xi_{ion}$ are vital for the determination of this matter. 

On the right side of Figure \ref{fig:q_test}, we test three different scenarios of $f_{esc}$ with $M_{UV, lim}=-17$. For the common assumption of $f_{\text{esc}}\sim0.2$, the epoch of reionisation is $z\sim12-7.5$ and considerably earlier than what observations indicate. In contrast, the higher abundance of source galaxies predicted by the EDE model and an efficient production rate ($\xi_{ion}$) makes it possible to reconstruct a consistent reionisation history with a smaller escape fraction, $f_{esc}\sim0.05-0.1$ which is also reflected in the $\chi^2$ values listed in Table \ref{tab:ede_chi2}. Therefore, the introduction of extra galaxies from EDE is capable of compensating for the ionising photon deficit that corresponds to the decline of both the escape fraction and the contribution from the faintest galaxies. As mentioned above, the high $f_{esc}$ is a tentative postulate that lacks observational determination. The more pragmatic assumption thus should be a slightly lower $f_{esc}\sim0.05-0.1$ due to their relatively younger stellar population but a ubiquitous absence of acute activities, e.g., AGN, to create efficient channels for higher $f_{esc}$ to take place, in which case the EDE-based prediction aligns with current data.

\begin{figure}
    \centering
    \includegraphics[width=\linewidth]{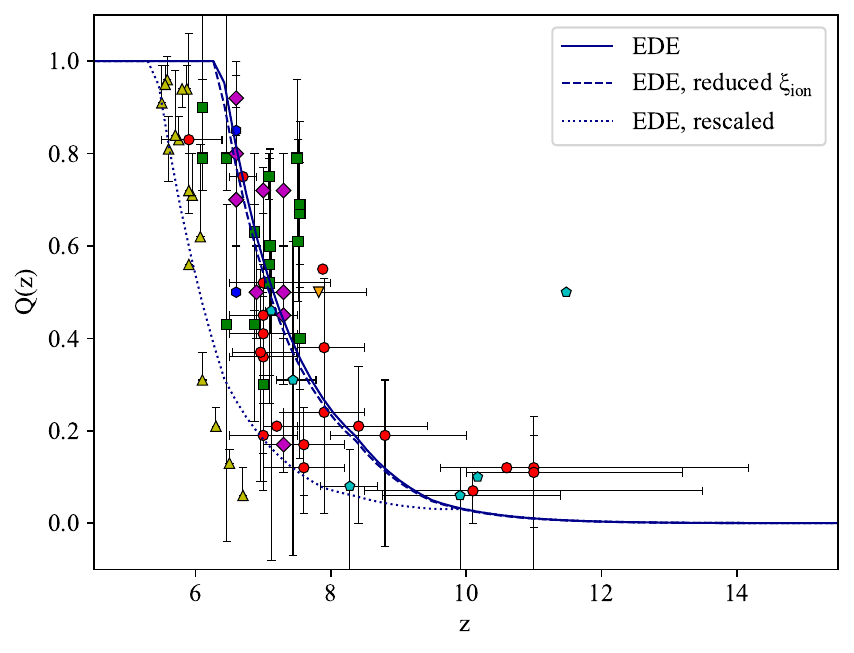}
    \caption{The reionisation history with two values of $\xi_{ion}$. 
    Given a higher $\xi_{ion}$, EDE predicts a fast rate of reionisation and reconstructs a consistent reionisation history with a less stringent parameter set ($(f_{esc}, \rm \lg{\xi_{ion}/Hz \ erg^{-1}}, M_{UV,lim})=(0.05, 25.8, -15)$). Alternatively, when $\rm \lg[\xi_{ion}/Hz \ erg^{-1}]$ is reduced to $25.3$ as is suggested by \cite{2024MNRAS.535.2998S}, generating a similar result requires an escalated $f_{esc}=0.15$ that marks the upper limit of $f_{esc}$ for EDE model, which is still lower than the previous assumption of $0.2$. On the other hand, given that the theoretical LFs are rescaled to match the data points with intermediate brightness, as in Figure \ref{fig:lf}, the insufficiency of the faint galaxies would greatly delay the entire reionisation process (dotted line). 
    }
    \label{fig:q_model}
\end{figure}

Generally speaking, our results indicate that the moderate choices of $f_{esc}=0.05\sim0.1$ and $M_{UV, lim}=-17\sim-15$ are sufficient for the EDE model to portray a cosmic reionisation history consistent with current observations. In particular, suppose we prioritise the completion time of EoR over its progression, the choice of these parameters can be further relaxed to $f_{esc}=0.05$ and $M_{UV, lim}=-17$ that optimally alleviate the reionisation photon budget deficit. Finally, in Figure \ref{fig:q_model} we compare the results of the reionisation history in the EDE model.
with two values of $\xi_{ion}$. In the higher case of $\rm \lg [\xi_{ion}/Hz \ erg^{-1}]=25.8$, the result is well aligned with the data on top of considerably relaxed requirements of $f_{esc}$ and $M_{UV, lim}$
Alternatively, when setting $\rm \lg [\xi_{ion}/Hz \ erg^{-1}]=25.3$ as is suggested by \cite{2024MNRAS.535.2998S}, the required $f_{esc}$ to generate a similar result is escalated to $0.15$, still lower than the previous assumption of $0.2$. The two cases thus mark the lower and upper limits of $f_{esc}$ being $0.05-0.15$ under the EDE model.
Hence, the EDE model could be further verified when more robust constraints on $\rm \lg [\xi_{ion}/Hz \ erg^{-1}]$ are available. Nonetheless, in either scenario, EDE could notably reduce the reliance on the high escape fraction or the extremely faint galaxies. Finally, we also plot the reionisation history corresponding to the rescaled LF in Figure \ref{fig:lf}. In this case, the entire reionisation process is greatly delayed due to the insufficiency of the faint galaxies (dotted line).

Therefore, we argue that a higher $\rho_{UV}$ under the EDE model that corresponds to extra ionising sources is necessary for providing sufficient ionising photons for reionisation with less dependence on a high average $f_{esc}$ and intensive contribution of the faintest galaxies.
Hence, if the key properties and distribution of the galaxies during EoR coincide with the extrapolations of their lower-redshift counterparts that constrain $f_{esc}\leq0.1$ and $M_{UV, lim}\leq-15$,
we conclude that it is legitimate to consider that an increment of the source galaxy abundance induced by the alteration of cosmology, i.e., EDE in this work, provides a coherent description of the current reionisation history observations, which in turn makes the proposition of using EoR to independently probe EDE, a promising theoretical model noted for its potential to alleviate the Hubble tension, yet primarily constrained by the CMB observations, a plausible contemplation.



\section{Summary} \label{sec:conclusion}

The application of the Early Dark Energy model provides an alternative perspective towards the alleviation of the Hubble tension. Despite the tentative observational preferences that emerged in some of the CMB and galaxy number density observations, a self-consistent cosmology model should also be able to coherently portray other observations, e.g., the reionisation history.
Therefore, in this work, we studied whether the EDE-based model could naturally reproduce an observationally consistent reionisation history with mitigated dependences on high $f_{esc}$ and faint $M_{UV, lim}$.

We first calculated the galaxy abundance in terms of stellar mass function (Figure \ref{fig:smf}) and luminosity function (Figure \ref{fig:lf}) under the EDE model.
Specifically, we proposed the hypothesis that the scaling relation of $M_h-M_\ast$ might exhibit limited sensitivity to the transition from $\Lambda$CDM to EDE. Although the initiation time of galaxy formation is expected to be earlier in the EDE model, the corresponding overall SFE might also be suppressed due to various heating effects and the earlier appearance of the feedback processes (Section \ref{sec:rho_uv}). Consequently, the competition of the two aspects may lead to the cancellation of the SFE variations during the EoR, i.e., the variation of the cosmological background exerts limited effects on the formation of individual galaxies. This effect could be particularly obvious for low mass systems, due to their shallower gravitational potentials. However, as the quantitative conclusion on this matter highly relies on the simulation of the galaxy formation under the EDE framework, we highlight this as a critical open question for future work and simplify our calculation by applying the pre-JWST $M_{h}-M_{\ast}$ relation to avoid the redundancy of using two independent approaches (JWST-enhanced SFE and alternative cosmology) simultaneously, which would otherwise lead to a double-fitting of the result.

Figure \ref{fig:smf} shows that the EDE-based result provides a good fit to the SMF data without the need for fine-tuning of the individual galaxy formation process embedded in the scaling relation. In particular, the consistency between data and the theoretical prediction at the low mass reflects a coherent depiction that they are the primary drivers of reionisation. Similarly, the LF in Figure \ref{fig:lf} also indicates the consistency between EDE prediction and observation, particularly at the massive and faint ends. Although the predicted galaxy abundance with intermediate brightness at $z\sim6, 9, 10$ is higher than the data, we attribute this to the result of the instantaneous SFR and dust attenuation. As mentioned above, the overestimation of the faint end proposed by \cite{2024MNRAS.533..860L} has been reduced with additional data from HST and JWST. Hence, a similar situation for the intermediate brightness may also be expected to take place with the piling up of more data in the future.


In order to verify the faint galaxy involvement, we tested the upper limit of the absolute magnitude by setting $M_{UV,lim}=-17, -15, -12$ and confirmed that the best fit requires $M_{UV,lim}\sim-15$ (left side of Figure \ref{fig:q_test}). However, considering our assumption of $\lg [\xi_{ion}/Hz \ erg^{-1}]=25.8$ is on the higher end of the current results, should the future observations yield a smaller value for this parameter, the involvement of the fainter galaxies may still be required. 
Next, we further tested three different escape fractions, $f_{esc}=0.05, 0.1, 0.2$ and found that the $0.1$ case demonstrates the best consistency with the data (right side of Figure \ref{fig:q_test}). Specifically, if we prioritise the completion time of EoR over its progression, the requirements of these parameters can be further relaxed to $f_{esc}=0.05$ and $M_{UV, lim}=-17$, simultaneously.

Finally, suppose the $f_{esc}$ is allowed to be higher at high redshift, we tested the requirement for this parameter with two choices of $\xi_{ion}$. As shown in Figure \ref{fig:q_model},
if $\xi_{ion}$ is not significantly greater than the pre-JWST result \citep[][]{2024MNRAS.535.2998S}, the required $f_{esc}=0.15$ is higher than the observational results at $z\leq4$, i.e., $f_{esc}\sim0.05$, yet still lower than the previous canonical value of $0.2$, suggesting that the reliance of $f_{esc}$ is reduced in either cases.
Furthermore, were the theoretical LFs to be rescaled to match the data points with intermediate brightness at $z\sim6-10$ (Figure \ref{fig:lf}), the resulting reionisation history would be greatly delayed, which reveals that the extra galaxies predicted by EDE are indispensable to the reconstruction of a coherent reionisation history.
Given these results, should the pillar of a considerably high $f_{esc}$ or $M_{UV, lim}$ be confirmed too frail to withstand the scrutiny of the observations to come, we argue that the EDE theory portrays a coherent cosmic reionisation history consistent with current observations, which, in turn, also makes the reionisation history a complementary probe of the EDE model, and the theory itself worthy of further deliberation.

\begin{acknowledgments}
We thank Hengjie Lin, Charlotte Mason, and Xiaolei Meng for constructive comments. We also thank the referee for the insightful suggestions, which greatly improved this manuscript. WL and HZ are supported by the National Key R\&D Program of China grant No. 2022YFF0503400 and by the China Manned Space Program through its Space Application System. 
XW is supported by the National Key R\&D Program of China No.2025YFF0510603, the National Natural Science Foundation of China (grant 12373009), the CAS Project for Young Scientists in Basic Research Grant No. YSBR-062, the China Manned Space Program, with grant no. CMS-CSST-2025-A06, and the Fundamental Research Funds for the Central Universities. XW acknowledges the support of the Xiaomi Young Talents Program and the work carried out, in part, at the Swinburne University of Technology, sponsored by the ACAMAR visiting fellowship.

\end{acknowledgments}

%

\vspace{5mm}
\facilities{HST, Spitzer/IRAC, JWST}



\software{\texttt{AxiCLASS} \citep{2018PhRvD..98h3525P,2020PhRvD.101f3523S}, \texttt{hmf} \citep{2021A&C....3600487M, 2013A&C.....3...23M}}





\bibliography{sample631}{}
\bibliographystyle{aasjournal}



\end{document}